\documentclass[journal]{IEEEtran}

\ifCLASSINFOpdf
\else
   \usepackage[dvips]{graphicx}
\fi
\usepackage{url}

\usepackage{graphicx}
\usepackage{ascmac}
\usepackage{amsmath,amssymb}
\usepackage{bm}
\usepackage{algorithmic}
\usepackage{algorithm}
\usepackage{cite}
\usepackage{subcaption}

\newcommand{\minimize}{\operatornamewithlimits{\mathrm{minimize}}}
\newcommand{\argmin}{\operatornamewithlimits{\mathrm{arg\ min}}}
\newcommand{\norm}[1]{\left\lVert#1\right\rVert}

\allowdisplaybreaks[1]
\begin{document}
\title{Depth-Aided Color Image Inpainting\\in Quaternion Domain}
\author{Shunki Tatsumi, Ryo Hayakawa, \IEEEmembership{Member, IEEE}, and Youji Iiguni, \IEEEmembership{Member, IEEE}
\vspace{-1mm}
\thanks{Shunki Tatsumi and Youji Iiguni are with the Graduate School of Engineering Science, Osaka University, 560-8531, Osaka, Japan (e-mail: tatsumi@sip.sys.es.osaka-u.ac.jp, iiguni@sys.es.osaka-u.ac.jp).}
\thanks{Ryo Hayakawa is with the Institute of Engineering, Tokyo University of Agriculture and Technology, 184-8588, Tokyo, Japan (e-mail: hayakawa@go.tuat.ac.jp).}
\thanks{© 2025 IEEE.  Personal use of this material is permitted.  Permission from IEEE must be obtained for all other uses, in any current or future media, including reprinting/republishing this material for advertising or promotional purposes, creating new collective works, for resale or redistribution to servers or lists, or reuse of any copyrighted component of this work in other works.}}
\markboth{ACCEPTED TO IEEE SINGAL PROCESSING LETTERS}
{Tatsumi \MakeLowercase{\textit{et al.}}: Depth-Aided Color Image Inpainting in Quaternion Domain}
\maketitle
\begin{abstract}
In this paper, we propose a depth-aided color image inpainting method in the quaternion domain, called depth-aided low-rank quaternion matrix completion (D-LRQMC). 
In conventional quaternion-based inpainting techniques, the color image is expressed as a quaternion matrix by using the three imaginary parts as the color channels, whereas the real part is set to zero and has no information. 
Our approach incorporates depth information as the real part of the quaternion representations, leveraging the correlation between color and depth to improve the result of inpainting. 
In the proposed method, we first restore the observed image with the conventional LRQMC and estimate the depth of the restored result. 
We then incorporate the estimated depth into the real part of the observed image and perform LRQMC again.
Simulation results demonstrate that the proposed D-LRQMC can improve restoration accuracy and visual quality for various images compared to the conventional LRQMC. 
These results suggest the effectiveness of the depth information for color image processing in quaternion domain. 
\end{abstract}
\begin{IEEEkeywords}
Image inpainting, quaternion, depth estimation, low-rank matrix approximation
\end{IEEEkeywords}
\IEEEpeerreviewmaketitle
\section{Introduction}
\IEEEPARstart{I}{mage} inpainting is one of fundamental problems in the field of image processing. 
The primary goal of inpainting is to restore the missing or damaged regions of the image in a visually plausible manner, seamlessly integrating them with the surrounding intact areas. 
This technique has wide-ranging applications, including the restoration of old photographs, removal of unwanted objects, and error concealment in transmission\cite{guillemot2013image}.

Low-rank approximation methods are powerful tools for image inpainting~\cite{candes2012exact, keshavan2010matrix,wang2014rank,lin2015accelerated,yu2019new, hu2012fast, shang2017bilinear}. 
These methods exploit the inherent low-rank structure of natural images, allowing the image matrix to be approximated by a matrix of significantly lower rank. 
In image inpainting, the restored image is usually computed by the low-rank matrix approximation under the constraint that no pixels other than the missing ones should be changed. 

One major problem of naive low-rank matrix approximation is that the matrix rank minimization problem is NP-hard~\cite{gu2017weighted,xie2016weighted,chen2017denoising,kang2015robust}. 
To address this issue, several surrogate functions of the matrix rank have been developed, such as the nuclear norm~\cite{candes2012exact, liu2012tensor}, Schatten $\gamma$-norm~\cite{frank1993statistical}, weighted nuclear norm~\cite{gu2017weighted}, and log-determinant penalty~\cite{kang2015logdet}. 
Minimizing such function under the pixel constraints is a popular approach for grayscale image inpainting.
Another method for low-rank matrix approximation is low-rank matrix factorization~\cite{liu2012tensor, chen2013simultaneous}. 
Low-rank matrix factorization aims to represent the target matrix as a product of two matrices whose sizes are smaller than that of the target matrix. 
In the context of image inpainting, we perform the matrix factorization under the pixel constraints. 

For the restoration of color images, various methods in the quaternion domain has been proposed. 
A quaternion is an extension of a complex number and is composed of one real part and three imaginary parts. 
By mapping the red, blue, and green channels of a color image to the three imaginary parts, respectively, a color image can be expressed as a quaternion matrix~\cite{pei1997novel}. 
The advantage of utilizing quaternions is that we can process all color channels simultaneously and use the correlation between each color channel~\cite{chen2014removing, subakan2011quaternion, chen2012quaternion, xu2015vector}. 
Quaternion is an effective representation of color images and have been used to various problems, such as denoising~\cite{gai2015denoising, chen2019low}, inpainting~\cite{chen2019low, miao2021color, jia2019robust}, edge detection~\cite{hu2018phase, xu2010color}, face recognition~\cite{zou2016quaternion, zou2019grayscale}, and video recovery~\cite{miao2020low}.
Color image inpainting methods using quaternions, such as low-rank quaternion approximation (LRQA)~\cite{chen2019low} and low-rank quaternion matrix completion (LRQMC)~\cite{miao2021color}, have shown improved accuracy compared to several methods in the real-valued domain.
In conventional color image representations using quaternions, however, the real part of the quaternion is usually set to zero and is not used as information for restoration. 

\begin{figure}[t]
    \centering
    \includegraphics[width=\columnwidth]{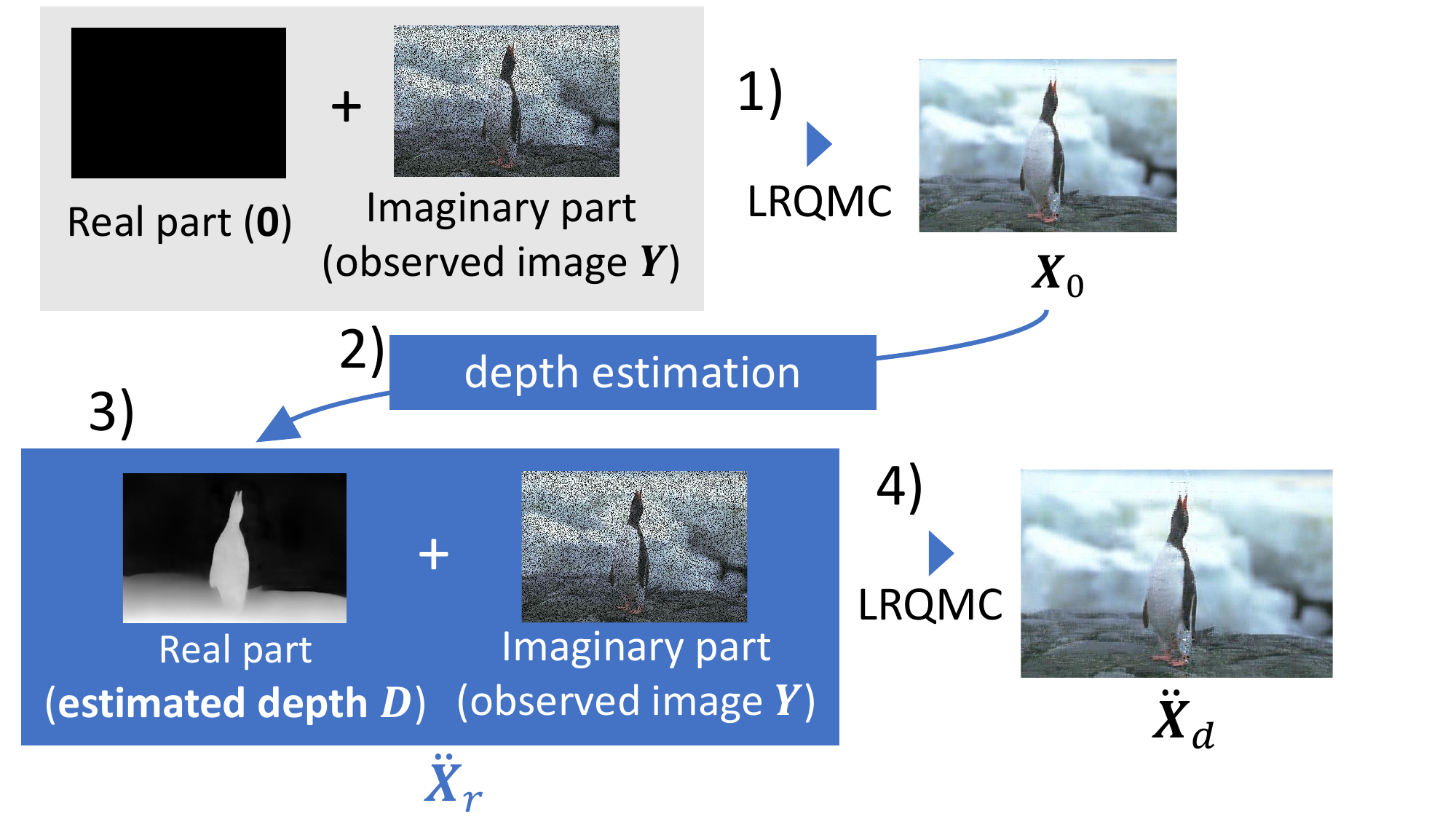}
    \caption{The flow of the proposed D-LRQMC. Each optimization step solves the problem~\eqref{optimization_problem}. The monocular depth estimation method~\cite{miangoleh2021boosting} is used for depth estimation.}
    \label{dlrqmc_flow}
\end{figure}

In this paper, we propose a color image inpainting method named depth-aided LRQMC (D-LRQMC) in Fig.~\ref{dlrqmc_flow}. 
The key idea of the proposed D-LRQMC is to incorporate depth information into the real part of the quaternion representation, thereby utilizing the correlation between depth and color to enhance inpainting performance. 
Here, depth refers to the distance from the camera to the object, and various methods have been developed to estimate the depth of each pixel from an image. 
By using the depth value as the real part, we can integrate depth information into various conventional inpainting techniques. 
A similar idea has also been used for face recognition in the quaternion domain recently~\cite{Shao2024-oa, Shao2024-yq}. 
In the proposed inpainting method, since depth cannot be correctly estimated from a missing image, we first restore the image with LRQMC, which is a promising conventional method in the quaternion domain. 
Depth estimation is then performed on the tentative restored image. 
For depth estimation, we use machine learning-based monocular depth estimation model~\cite{miangoleh2021boosting, alhashim2018high}. 
Finally, we put the result of the depth estimation into the real part of the observed image and perform LRQMC again. 
Experimental results shows that 93\% of the test images were restored with improved accuracy.
These results suggest the effectiveness of the proposed approach, i.e., the use of the depth information, for color image processing in quaternion domain. 
\section{Color Image Inpainting} \label{formulation}
In this chapter, we formulate the inpainting problem.
The observed color image $\bm{Y} \in \mathbb{R} ^ {M \times N \times 3}$ with missing pixels can be expressed as
\begin{align}
\bm{Y} = \mathcal{P}_{\bm{B}} (\bm{X}),
\end{align}
where $\bm{X} \in \mathbb{R} ^ {M \times N \times 3}$ is the original color image and $\bm{B} \in \{0, 1\} ^ {M \times N}$ is the binary mask. 
$\mathcal{P}_{\bm{B}} (\cdot)$ is the projection onto the linear space of matrices supported on $\bm{B}$, i.e., the $(m, n, c)$ element of $\mathcal{P}_{\bm{B}}(\bm{X})$ is given by
\begin{align}
(\mathcal{P}_{\bm{B}}(\bm{X}))_{m, n, c} = 
    \begin{cases}
        x_{m, n, c}\quad &\text{if } b_{m, n} = 1\\
        0\quad &\text{if } b_{m, n} = 0
    \end{cases},
\end{align}
where $x_{m, n, c}$ is the $(m, n, c)$ element of $\bm{X}$ and $b_{m, n}$ is the $(m, n)$ element of $\bm{B}$. 
The ranges of $m, n$ are  $m = 1, \ldots ,M$, $n = 1, \ldots ,N$, and $c=1, 2, 3$ corresponds to red, green, and blue, respectively.
\section{Conventional Methods} \label{conventional}
\subsection{Image Inpainting in Real Domain}
The minimization of nuclear norm, i.e., sum of singular values, is a powerful technique for inpainting using the idea of matrix rank minimization. 
This is because the nuclear norm is known to be an effective convex function for evaluating the low-rank nature of matrices~\cite{candes2012exact}. 
In this approach, we aim to minimize the nuclear norm of the matrix by varying only the missing parts of the target matrix.

Another approach is matrix factorization via alternating minimization~\cite{jain2013low}. 
In this method, the target low-rank matrix $\bar{\bm{X}} \in \mathbb{R}^{M \times N}$ is approximated by the bi-linear form as $\bar{\bm{X}} \approx \bm{U} \bm{V}$, where $\bm{U} \in \mathbb{R} ^{M \times K}$ and $\bm{V} \in \mathbb{R} ^{K \times N}$. 
In this form, the rank of the product of the matrices $\bm{UV}$ is at most $K$. 
By computing $\bm{UV}$ closer to the target matrix $\bar{\bm{X}}$ while satisfying the constraints, we obtain a restored image that can be approximated by a matrix of rank at most $K$.
\subsection{Image Inpainting in Quaternion Domain}
Quaternion is an extension of complex numbers. 
A quaternion $\ddot{q} \in \mathbb{H}$ has one real part and three imaginary parts, where $\mathbb{H}$ is the set of all quaternions. 
Specifically, it is represented as
\begin{align}
    \ddot{q} = a + bi + cj + dk,
\end{align}
where $a, b, c, d\in \mathbb{R}$ are real numbers and $i, j, k$ are imaginary units, which obey the quaternion rules $i^2 = j^2 = k^2 = ijk = -1$, $ij = -ji = k$, $jk = -kj = i$, and $ki = -ik = j$. 

By incorporating the RGB components of a color image pixel into the three imaginary parts, each pixel of a color image can be represented by a quaternion, i.e.,
\begin{align}
    \ddot{q}_{m,n} = r_{m, n}i + g_{m, n}j + b_{m, n}k, \label{eq:quaternion_color}
\end{align}
where $r_{m, n}$, $g_{m, n}$, and $b_{m, n}$ are the red, green, and blue components corresponding to the pixel at position $(m, n)$ in the color image, respectively.
Thus, the entire color image can be represented by a quaternion matrix.

Various quaternion-based matrix completion method can be viewed as an extension of real-valued matrix completion to the quaternion domain. 
In LRQMC~\cite{miao2021color}, for example, the optimization problem is given by
\begin{align}
    \label{optimization_problem}
    \minimize_{\substack{f(\ddot{\bm U})\in \mathbb{C}^{2M \times 2K}, \\f(\ddot{\bm V})\in \mathbb{C}^{2K \times 2N}, \\ \ddot{\bm X}\in \mathbb{H}^{M \times N}}} \quad 
    &\frac{1}{2} \norm{f(\ddot{\bm U})f(\ddot{\bm V})-f(\ddot{\bm X})}^2_\text{F} \notag \\
    &+
    \frac{\lambda}{2} \left( \norm{f(\ddot{\bm U})}^2_\text{F} + \norm{f(\ddot{\bm V})}^2_\text{F} \right)\notag \\
    \mathrm{subject\ to}\quad &\mathcal{P}_{\bm{B}} (\ddot{\bm X} - \ddot{\bm{Y}}) = \bm{O},
\end{align}
where $\ddot{\bm Y} \in \mathbb{H}^{M \times N}$ is the quaternion matrix corresponding to the observed color image, $\bm{O}$ is the matrix whose elements are all zero, and $\lambda$ ($>0$) and positive integer $K$ are parameters. 
The operator $f : \mathbb{H}^{M \times N} \to \mathbb{C}^{2M \times 2N}$ is defined as 
\begin{align}
    \label{f}
    f(\ddot{\bm Q}) = 
    \begin{pmatrix}
        \bm{Q}_a & \bm{Q}_b \\
        -\bm{Q}_b^\ast & \bm{Q}_a^\ast \\
    \end{pmatrix},
\end{align}
where $(\cdot)^{\ast}$ denotes the complex conjugate, $\ddot{\bm Q} = \bm{Q}_0 + \bm{Q}_1i+\bm{Q}_2j+\bm{Q}_3k = \bm{Q}_a + \bm{Q}_bj$, and $\bm{Q}_a = \bm{Q}_0 + \bm{Q}_1 i, \quad \bm{Q}_b = \bm{Q}_2 + \bm{Q}_3 i$. 
Hence, $f(\ddot{\bm Q})$ is uniquely computed from $\ddot{\bm Q}$. 
This function $f$ makes it possible to treat quaternions as complex numbers in the optimization.
$\norm{\cdot}_{\text{F}}$ denotes the Frobenius norm. 

Although the problem~\eqref{optimization_problem} is non-convex itself, it is convex with respect to each variable of $f(\ddot{\bm U})$, $f(\ddot{\bm V})$, and $\ddot{\bm X}$. 
Hence, in LRQMC, we solve the optimization problem iteratively by using an alternating minimization approach. 
Letting $\mathcal{G}(f(\ddot{\bm{U}}), f(\ddot{\bm{V}}), \ddot{\bm{X}})
= \frac{1}{2}\|f(\ddot{\bm U})f(\ddot{\bm V})-f(\ddot{\bm X})\|^2_\text{F} 
 + \frac{\lambda}{2}(\|f(\ddot{\bm U})\|^2_\text{F} + \|f(\ddot{\bm V})\|^2_\text{F})$, we can write the update equation as
\begin{subequations}
    \begin{align}
        f(\ddot{\bm{U}})^{t + 1} =& \argmin_{f(\ddot{\bm{U}})\in \mathbb{C}^{2M \times 2K}}  \mathcal{G}(f(\ddot{\bm{U}}), f(\ddot{\bm{V}})^t, \ddot{\bm{X}}^t), \\
        f(\ddot{\bm{V}})^{t + 1} =& \argmin_{f(\ddot{\bm{V}})\in \mathbb{C}^{2K \times 2N}}  \mathcal{G}(f(\ddot{\bm{U}})^{t+1}, f(\ddot{\bm{V}}), \ddot{\bm{X}}^t) ,\\
        \ddot{\bm{X}}^{t + 1} =& \argmin_{\ddot{\bm{X}} \in \mathbb{H}^{M \times N}}  \mathcal{G}(f(\ddot{\bm{U}})^{t+1}, f(\ddot{\bm{V}})^{t+1}, \ddot{\bm{X}}),\\
        &\mathrm{subject\ to}\quad \mathcal{P}_{\bm{B}} (\ddot{\bm X} - \ddot{\bm{Y}}) = \bm{O}.
    \end{align}
\end{subequations}

First, we consider the update of $f(\ddot{\bm{U}})$ and $f(\ddot{\bm{V}})$. 
By performing Wirtinger's derivative of $\mathcal{G}$ with respect to $f(\ddot{\bm{U}})$ and $f(\ddot{\bm{V}})$, the update equations can be written as 
\begin{subequations}
\label{fu=}
\begin{align}
    f(\ddot{\bm{U}})^{t+1} &= f(\ddot{\bm{X}})^t (f(\ddot{\bm{V}})^t)^\mathsf{H} \bm{\Psi_{\ddot{\bm{V}}}}^t, \label{U} \\
    f(\ddot{\bm{V}})^{t+1} &= \bm{\Phi_{\ddot{\bm{U}}}}^t(f(\ddot{\bm{U}})^{t+1})^\mathsf{H} f(\ddot{\bm{X}})^t, \label{V}
\end{align}
\end{subequations}
respectively. 
Here, $\bm{\Psi_{\ddot{\bm{V}}}}^t$ and $\bm{\Phi_{\ddot{\bm{U}}}}^t$ are given by $\bm{\Psi_{\ddot{\bm{V}}}}^t = \left( f(\ddot{\bm{V}})^t (f(\ddot{\bm{V}})^t)^\mathsf{H} + \lambda \bm{I}_{2K}  \right)^\dag$ and $\bm{\Phi_{\ddot{\bm{U}}}}^t = \left( (f(\ddot{\bm{U}})^{t+1})^\mathsf{H} f(\ddot{\bm{U}})^{t+1}+ \lambda \bm{I}_{2K}  \right) ^\dag$, respectively, where $(\cdot)^{\mathsf{H}}$ represents the Hermitian transpose and $(\cdot)^\dag$ denotes the pseudo-inverse matrix.

Given that the constraint $\mathcal{P}_{\bm{B}} (\ddot{\bm X} - \ddot{\bm{Y}}) = \bm{O}$, in the update of $\ddot{\bm{X}}$, only the elements of $\ddot{\bm{X}}$ corresponding to $b_{m, n} = 0$ can be changed. 
From~\eqref{optimization_problem}, we can see that the optimal solution is the matrix whose $(m, n)$ element with $b_{m,n} = 0$ is equal to the corresponding element of $f(\ddot{\bm{U}})f(\ddot{\bm{V}})$. 
Therefore, the result is obtained by
\begin{align}
  \ddot{\bm{X}}^{t + 1} = \mathcal{P}_{\bm{B}^C} \left( f^{-1}(f(\ddot{\bm{U}}^{t + 1}) f(\ddot{\bm{V}}^{t + 1})) \right) + \ddot{\bm{Y}}, \label{X}
\end{align}
where $\bm{B}^C$ is the complement of $\bm{B}$ and $f^{-1}$ denotes the inverse operation of $f$. 

The algorithm of LRQMC is summarized in Algorithm~\ref{algo1}. 
The rank estimation process in LRQMC is omitted for simplicity. 
\begin{algorithm}[t]
\caption{LRQMC~\cite{miao2021color}}
    \label{algo1}
    \begin{algorithmic}[1]
        \REQUIRE observed image $\ddot{\bm{Y}}\in \mathbb{H}^{M \times N}$
        \ENSURE $\ddot{\bm{X}}^{t}$
        \STATE $t  \leftarrow 0, \ddot{\bm{X}}^{0} \leftarrow \ddot{\bm{Y}}$
        \STATE initialize $f(\ddot{\bm{V}})^0 \in \mathbb{C}^{2K \times 2N}$ randomly
        \REPEAT
        \STATE $\bm{\Psi_{\ddot{\bm{V}}}}^t \leftarrow \left( f(\ddot{\bm{V}})^t (f(\ddot{\bm{V}})^t)^\mathsf{H} + \lambda \bm{I}_{2K}  \right) ^\dag$
        \STATE $f(\ddot{\bm{U}})^{t + 1} \leftarrow f(\ddot{\bm{X}})^t (f(\ddot{\bm{V}})^t)^\mathsf{H} \bm{\Psi_{\ddot{\bm{V}}}}^t$
        \STATE $\bm{\Phi_{\ddot{\bm{U}}}}^t \leftarrow \left( (f(\ddot{\bm{U}})^{t+1})^\mathsf{H} f(\ddot{\bm{U}})^{t+1}+ \lambda \bm{I}_{2K}  \right) ^\dag$
        \STATE $f(\ddot{\bm{V}})^{t + 1} \leftarrow \bm{\Phi_{\ddot{\bm{U}}}}^t(f(\ddot{\bm{U}})^{t+1})^\mathsf{H} f(\ddot{\bm{X}})^t$
        \STATE $\ddot{\bm{X}}^{t + 1} \leftarrow \mathcal{P}_{\bm{B}^C} \left( f^{-1}(f(\ddot{\bm{U}}^{t + 1}) f(\ddot{\bm{V}}^{t + 1})) \right) + \ddot{\bm{Y}}$
        \STATE $t \leftarrow t+1$
        \UNTIL {$\mathbf{convergence}$}
    \end{algorithmic} 
\end{algorithm}

\section{Proposed Method} \label{proposed}
In this section, we describe the proposed D-LRQMC, which is a depth-aided inpainting method based on LRQMC. 
The proposed method incorporates depth information into the real part of the quaternions.
Specifically, our key idea is to consider the quaternion
\begin{align}
    \ddot{q}_{m,n} = d_{m, n} + r_{m, n}i + g_{m, n}j + b_{m, n}k
\end{align}
instead of~\eqref{eq:quaternion_color}, 
where $d_{m, n} \in \mathbb{R}$ denotes the depth of the pixel $(m, n)$. 
By utilizing the correlation between color and depth in natural images, we aim to improve the restoration accuracy. 

The flow of the proposed method is shown in Fig.~\ref{dlrqmc_flow}.
Each step of the proposed method is as follows:
\begin{enumerate}
  \item Since depth cannot be correctly estimated from images with many missing pixels, we first restore the image with the conventional LRQMC. 
    We solve the optimization problem in~\eqref{optimization_problem} for observed image $\bm{Y}$ and obtain output image $\bm{X}_0$. 
  \item We estimate the depth map $\bm{D} \in [0, 255]^{M \times N}$ from the tentative result $\bm{X}_{0}$ obtained by LRQMC. 
    We first express $\bm{X}_0$ as a color image and then obtain the grayscale depth map $\bm{D}$ from the color image. 
    For depth estimation, we use the machine learning-based model~\cite{miangoleh2021boosting} that can estimate depth from a single color image in this paper. 
  \item We define a quaternion matrix $\ddot{\bm{X}}_r$ composed of the depth map $\bm{D}$ as the real part and the observation matrix $\bm{Y}$ as the imaginary part. 
    Since the real part of $\bm{Y}$ and $\bm{D}$ have the same size, we can insert the pixel values of $\bm{D}$ directly into the real part of $\ddot{\bm{X}}_r$.
  \item Perform LRQMC again for $\ddot{\bm{X}}_r$ and get the final output $\ddot{\bm{X}}_d$.
\end{enumerate}

Due to the additional steps of depth estimation and the second execution of LRQMC, the computational complexity of the proposed method increases compared to the conventional LRQMC with $\mathcal{O}(KMN)$~\cite{miao2021color}. 
It should also be noted that our approach can be applied to any inpainting methods in the quaternion domain, though we consider LRQMC as an example of promising methods in this paper. 
\begin{figure*}[t]
    \centering
    \begin{minipage}[t]{0.45\textwidth}
        \centering
        \includegraphics[width=\textwidth]{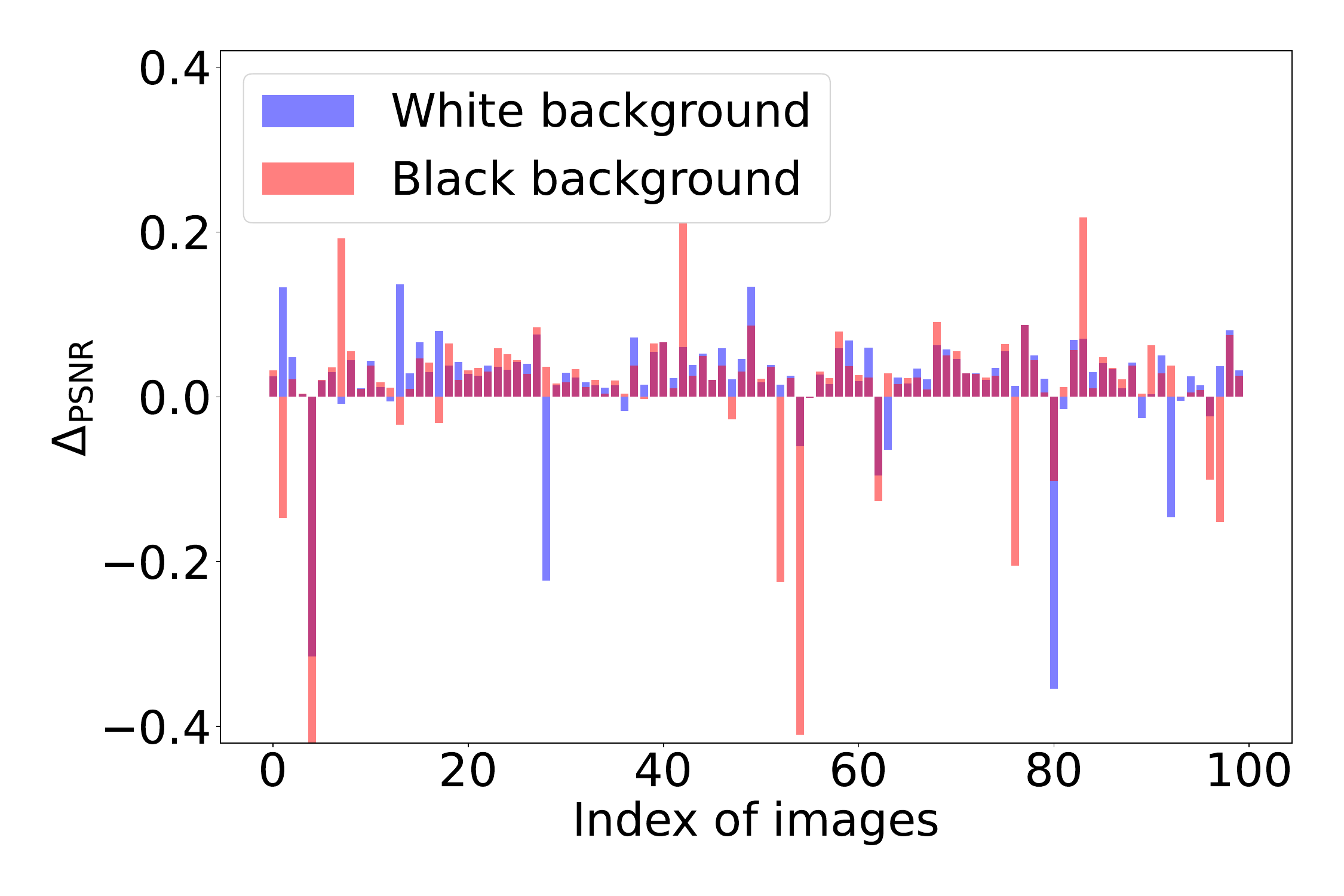}
        \vspace{-8mm}
        \subcaption{$\Delta_\text{PSNR}$}
        \label{S_PSNR_graph}
    \end{minipage}
    \hspace{5mm}
    \begin{minipage}[t]{0.45\textwidth}
        \centering
        \includegraphics[width=\textwidth]{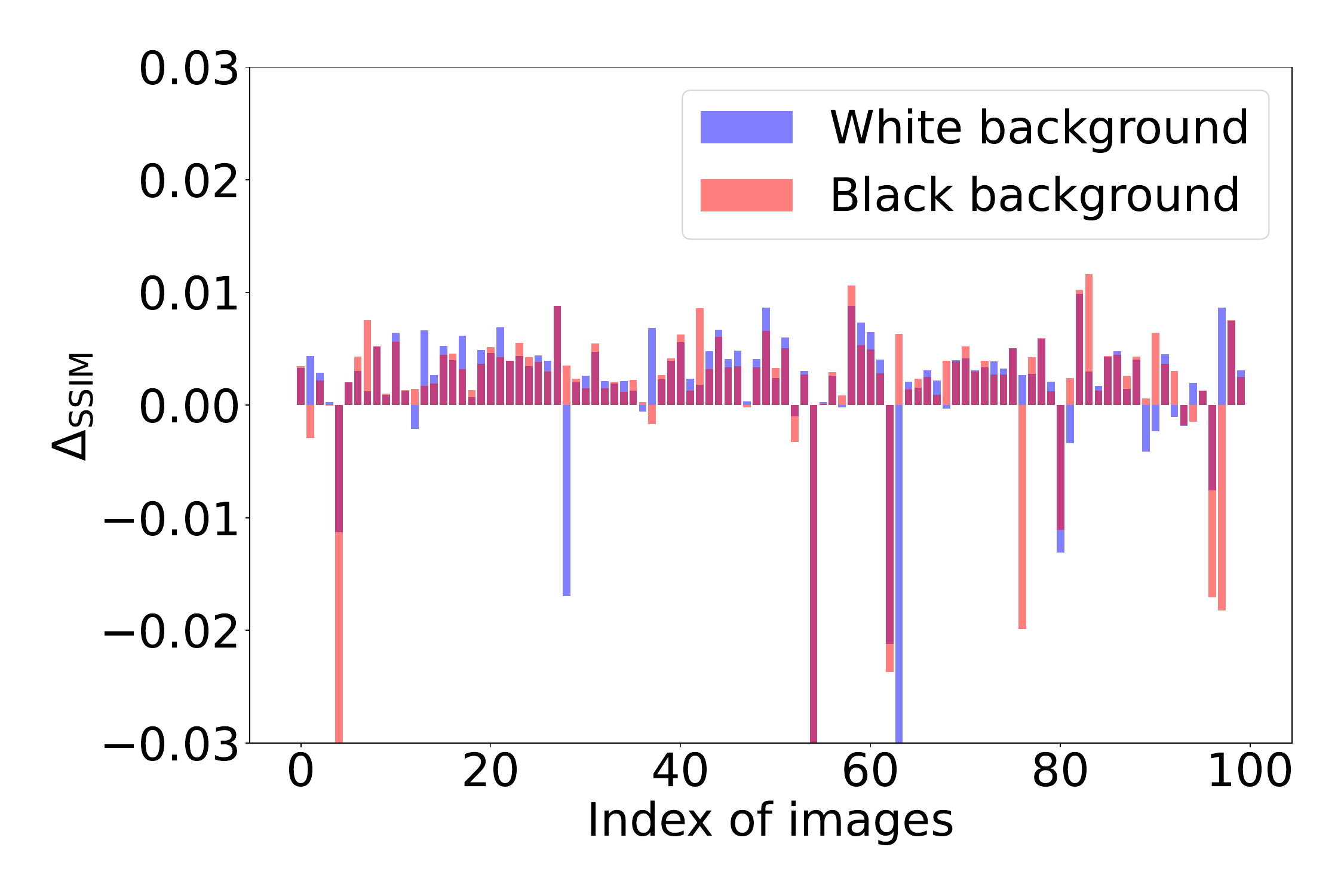}
        \vspace{-8mm}
        \subcaption{$\Delta_\text{SSIM}$}
        \label{S_SSIM_graph}
    \end{minipage}
    \vspace{-2mm}
    \caption{$\Delta_\text{PSNR}$ and $\Delta_\text{SSIM}$ for different depth map patterns: White background (blue) and Black background (red).}
    \label{fig:comparison}
\end{figure*}
\begin{figure*}[t]
    \centering
    \includegraphics[width=0.45\textwidth]{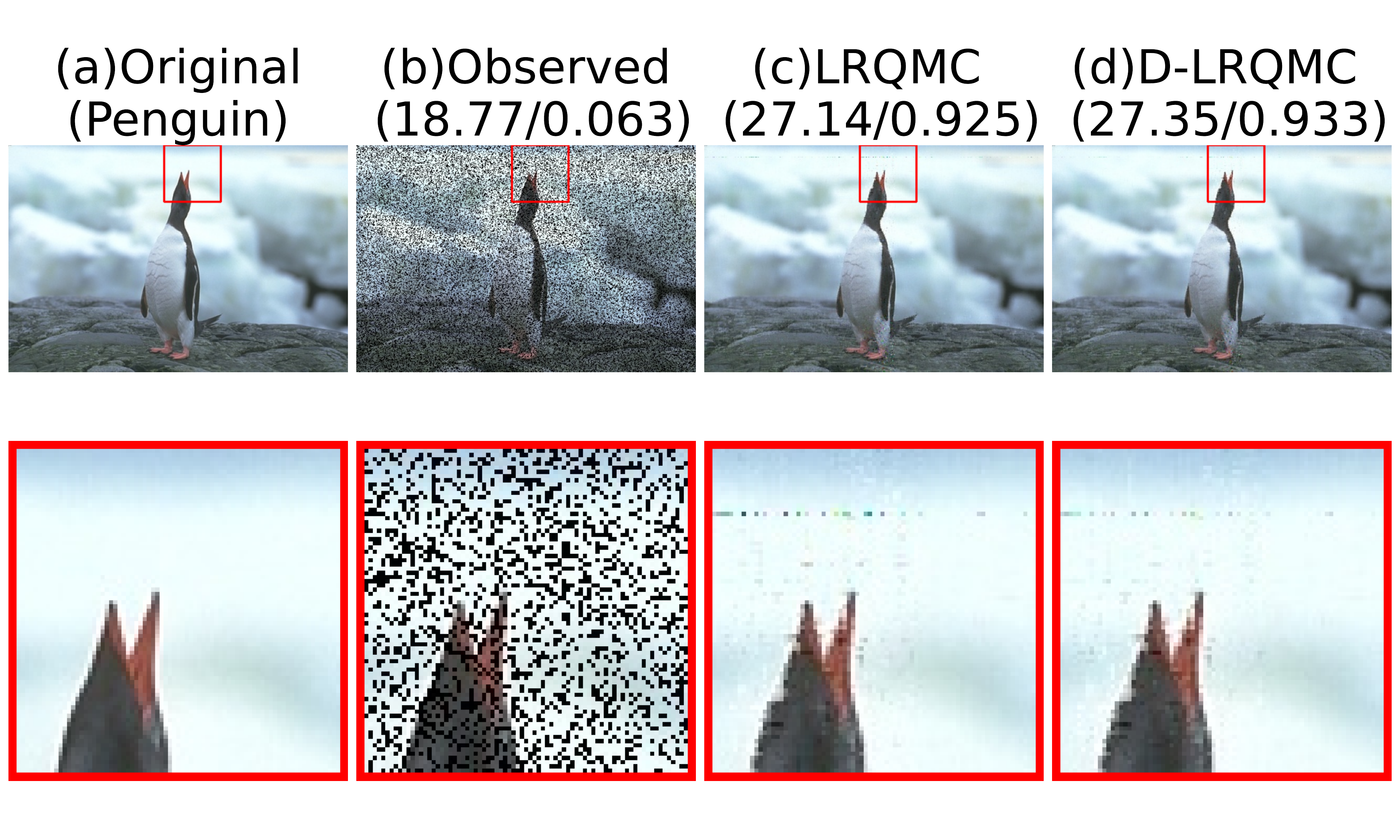}
    \hspace{5mm}
    \includegraphics[width=0.45\textwidth]{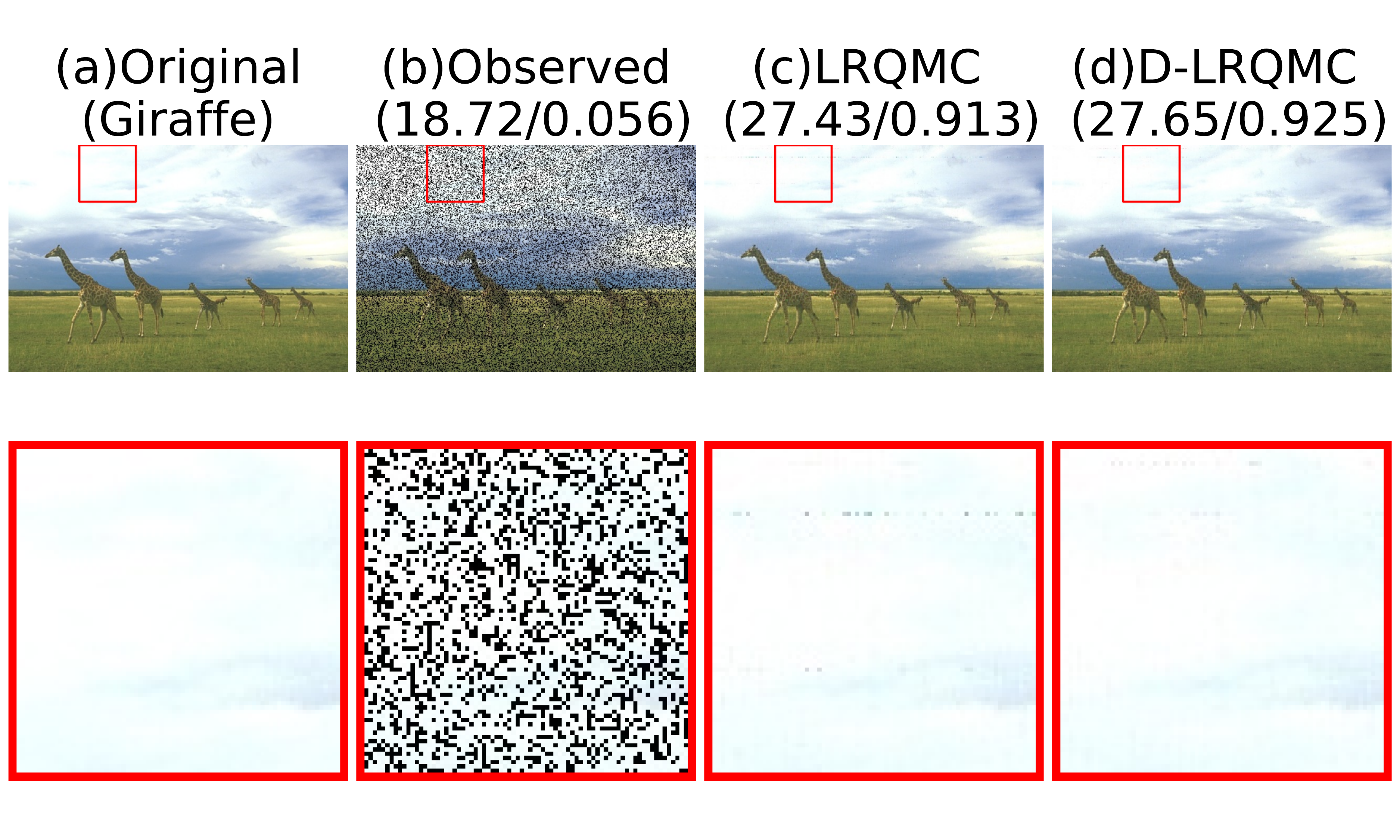}
    \vspace{-4mm}
    \caption{Images of the restoration results of two images (Penguin and Giraffe). (a) Original image (b) Observed image with 30\% of all pixels randomly missing (c) LRQMC~\cite{miao2021color} (d) D-LRQMC (Ours). The numbers in (b), (c), and (d) are the PSNR (dB) and SSIM of the respective images.}
    \label{result_image}
\end{figure*}
\section{Simulation Results and Discussion} \label{experiments}
We compare the performance of the proposed D-LRQMC with the conventional LRQMC~\cite{miao2021color} via computer simulations. 
Note that LRQMC outperforms many optimization-based inpainting methods~\cite{miao2021color}. 
The simulation is conducted by using a computer with Intel Core i9-10920X, 32~GB memory, and NVIDIA GeForce RTX 2060 SUPER. 

In the simulations, we use the Berkeley segmentation dataset\cite{bsds_dataset}. 
There are 100 clean color images of size $481 \times 321 \times 3$ in the whole dataset.
The observed images are obtained by randomly masking 30\% of the pixels in the original images. 
For both methods, we use the parameter $\lambda=1$ in~\eqref{optimization_problem} as in~\cite{miao2021color}.
The parameter $K$ for the matrix decomposition is fixed to $80$, which achieves good reconstruction performance for the conventional LRQMC. 
In both methods, each element of the initial value $\ddot{\bm{V}}^0$ in LRQMC follows the uniform distribution on $[0, 255]$. 
All other conditions are identical for comparison because the purpose of the simulation is to evaluate the improvement of the accuracy due to depth information. 

The performance of the inpainting was evaluated by peak signal to noise ratio (PSNR) in dB and structural similarity index measure (SSIM). 
To simplify the comparison, we define $\Delta_\text{PSNR} = \text{PSNR}_\text{D-LRQMC} - \text{PSNR}_\text{LRQMC}$ and $\Delta_\text{SSIM} = \text{SSIM}_\text{D-LRQMC} - \text{SSIM}_\text{LRQMC}$, where $\text{PSNR}_\text{D-LRQMC}$ and $\text{PSNR}_\text{LRQMC}$ (resp. $\text{SSIM}_\text{D-LRQMC}$ and $\text{SSIM}_\text{LRQMC}$) are the PSNR (resp. SSIM) values of results of LRQMC and D-LRQMC, respectively. 
These scores indicate how much better D-LRQMC is in terms of the accuracy compared to LRQMC. 
For the depth estimation in the proposed D-LRQMC, we consider two patterns of the depth map: 
White background (Far objects appear brighter and their pixel value is close to 255) and Black background (Far objects appear darker and pixel values are close to zero). 

The scores $\Delta_\text{PSNR}$ and $\Delta_\text{SSIM}$ for all images in the two patterns are shown in Fig.~\ref{fig:comparison}. 
Fig.~\ref{fig:comparison}(a) shows that 79\% of the images have a positive $\Delta_\text{PSNR}$ regardless of the depth map type.
Similarly, Fig.~\ref{fig:comparison}(b) indicates that 77\% of the images have a positive $\Delta_\text{SSIM}$ regardless of the depth map type. 
By selecting an appropriate depth map, the percentage increases to 93\%. 
We can see that the accuracy can be improved for many images by using the depth information as the real part of the quaternion matrix. 
Simulation results for different missing ratios can be found in the Supplementary Material. 

Fig.~\ref{result_image} presents the restoration results for two images. 
We show (a) original image, (b) observed image with 30\% missing pixels, (c) restoration result by the conventional LRQMC, and (d) restoration result by the proposed D-LRQMC. 
The numbers on each image represent the PSNR (dB) and SSIM values. 
The images below each set are magnified sections of the corresponding images to highlight the restoration quality.
This figure visually demonstrates the superior restoration accuracy of the proposed method. 
A possible reason for the improvement in the solid background areas is that the depth remains largely unchanged in those regions. 

In the proposed method, the average computation times per image for the first LRQMC, depth estimation, and second LRQMC were 1.79 s, 98.89 s, and 1.86 s, respectively. 
The computation time for the depth estimation was much longer than the inpainting process in our environment. 
\section{Conclusion}
\label{conclusion}
In this study, we have proposed a depth-aided color image inpainting method in the quaternion domain named D-LRQMC. 
The proposed method extends the conventional LRQMC by incorporating depth information as the real part of the quaternion representation, thereby leveraging the correlation between color and depth to enhance the restoration accuracy. 
Experimental results demonstrated that the proposed D-LRQMC method outperforms the conventional LRQMC for various images. 
For example, the use of depth information improves the SSIM for 93\% of the test images when an appropriate depth map is utilized. 
The visual quality of the restored images also confirmed the effectiveness of our approach. 
These findings confirm the effectiveness of utilizing depth information for quaternion-based image inpainting. 

Future work includes establishing criteria for selecting the appropriate depth pattern, potentially by leveraging blind image quality assessments~\cite{Moorthy2011-fd} to evaluate inpainting results. 
Additionally, applying the proposed depth-aided approach to more recent methods~\cite{Jia2022-hp,Miao2023-id,Xu2023-bw,Wu2024-ym,Miao2024-fp} and investigating the causes of extremely poor accuracy in certain images remain an important research direction. 
\bibliographystyle{IEEEtran}
\bibliography{reference}
\newpage
\appendices
\section{Supplementary Simulation Results}
In this section, we present additional simulation results to further evaluate the performance of the proposed D-LRQMC for different missing ratios. 
The other simulation settings remain the same as those used in the main paper.

\begin{figure}[!b]
    \centering
    \begin{minipage}[b]{0.48\textwidth}
        \centering
        \includegraphics[width=\textwidth]{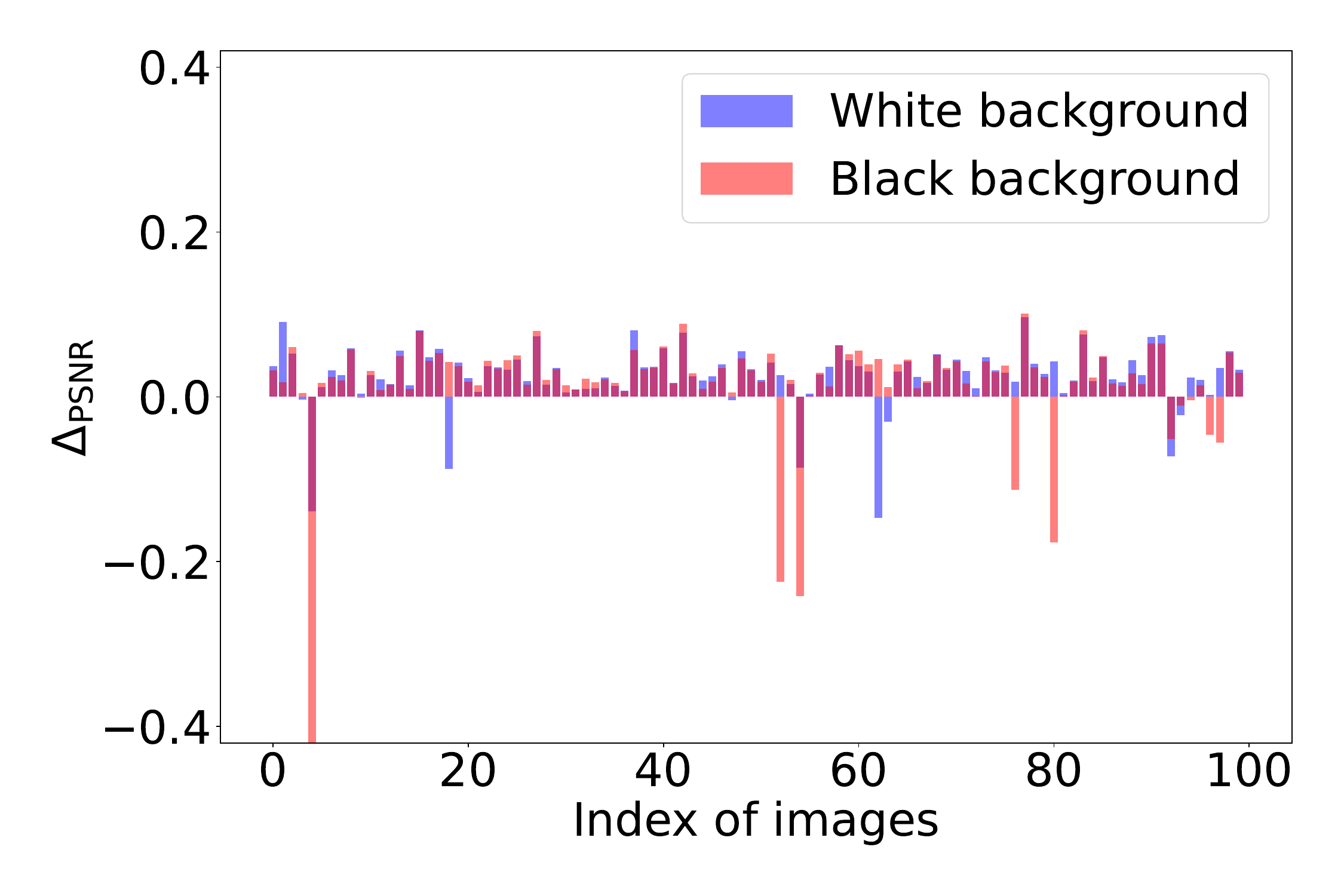}
        \subcaption{$\Delta_\text{PSNR}$}
        \label{S_PSNR_graph10}
    \end{minipage}
    \hfill
    \begin{minipage}[b]{0.48\textwidth}
        \centering
        \includegraphics[width=\textwidth]{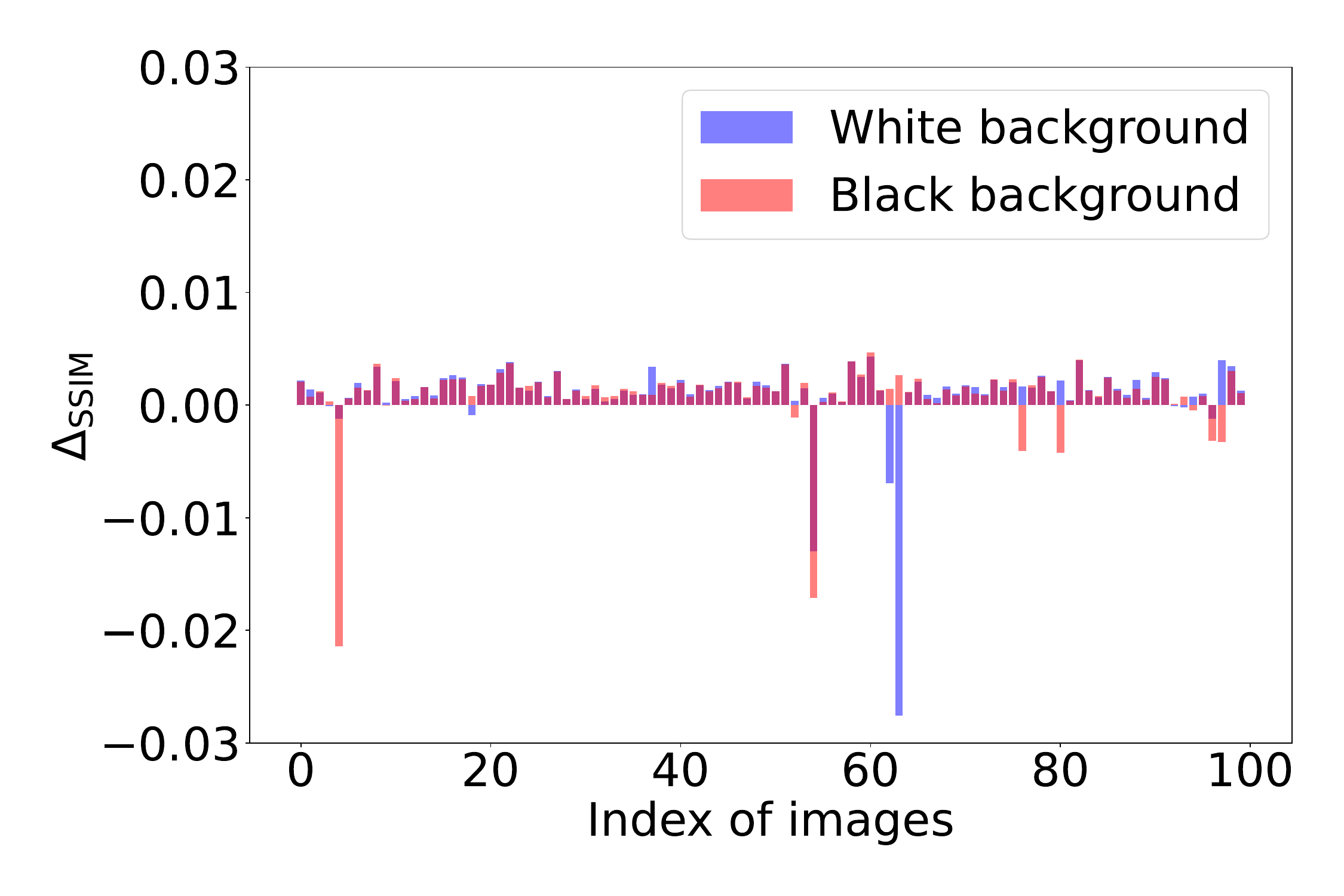}
        \subcaption{$\Delta_\text{SSIM}$}
        \label{S_SSIM_graph10}
    \end{minipage}
    \caption{$\Delta_\text{PSNR}$ and $\Delta_\text{SSIM}$ for different depth map patterns (masked pixels: 10\%).}
    \label{fig:comparison_10}
\end{figure}

Fig.~\ref{fig:comparison_10} shows the performance improvement $\Delta_\text{PSNR}, \Delta_\text{SSIM}$ when the observed image is obtained by randomly masking 10\% pixels of the original image. 
84\% of the images have a positive $\Delta_\text{PSNR}$ regardless of the depth map type, and the percentage increases to 96\% if we can select the appropriate depth map. 

\begin{figure}[!b]
    \centering
    \begin{minipage}[b]{0.48\textwidth}
        \centering
        \includegraphics[width=\textwidth]{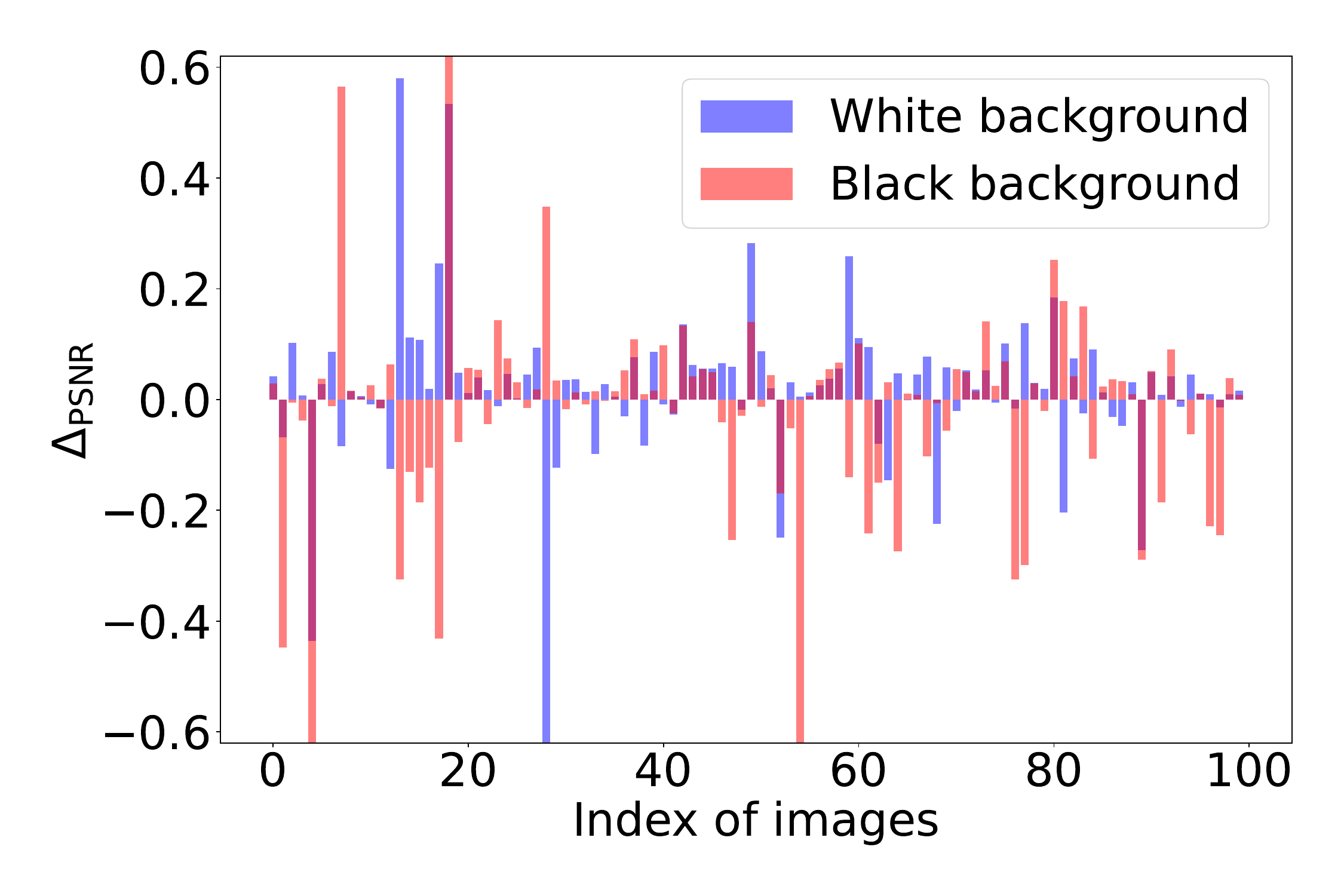}
        \subcaption{$\Delta_\text{PSNR}$}
        \label{S_PSNR_graph50}
    \end{minipage}
    \hfill
    \begin{minipage}[b]{0.48\textwidth}
        \centering
        \includegraphics[width=\textwidth]{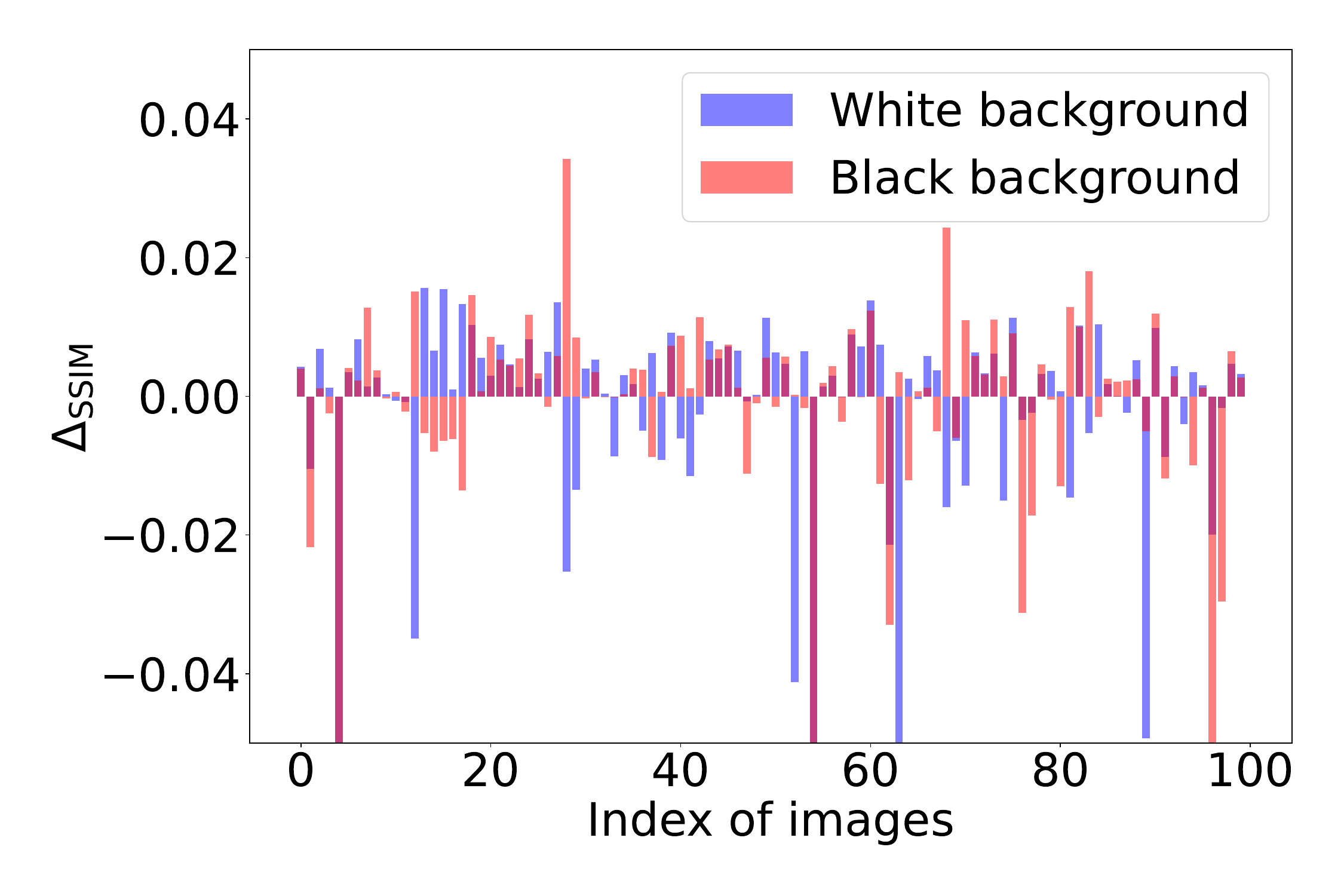}
        \subcaption{$\Delta_\text{SSIM}$}
        \label{S_SSIM_graph50}
    \end{minipage}
    \caption{$\Delta_\text{PSNR}$ and $\Delta_\text{SSIM}$ for different depth map patterns (masked pixels: 50\%).}
    \label{fig:comparison_50}
\end{figure}

Fig.~\ref{fig:comparison_50} shows the result for the observed images with 50\% missing pixels. 
The performance in this case is highly dependent on the type of depth map utilized. 
However, by selecting an appropriate depth map type, performance improvements are observed in 88\% of the images. 
A plausible explanation for this phenomenon is that the quality of the depth map generated from the output of LRQMC deteriorates as the proportion of missing pixels increases. 

From the results for various missing ratios, we can see that the performance improvement by the proposed approach depends on the problem settings and the performance of the original LRQMC. 
These findings suggest that while depth information can enhance inpainting quality, its effectiveness may vary depending on the proportion of the missing regions and the accuracy of the depth estimation. 
\end{document}